\documentclass[onecolumn,prd,amssymb,amsmath,preprintnumbers,aps,nofootinbib,superscriptaddress,12pt]{revtex4-2}
\usepackage[a4paper, portrait, margin=0.7in]{geometry}
\usepackage[utf8]{inputenc}
\usepackage{multirow}
\usepackage{lineno}
\usepackage{bm,color}
\usepackage[dvipsnames]{xcolor}
\usepackage{slashed} 
\usepackage{graphicx}
\usepackage{setspace}
\usepackage{multirow,booktabs,colortbl}
\usepackage{soul} 
\usepackage{multirow}
\usepackage{comment}
\usepackage{epstopdf} 
\usepackage{mathtools}
\usepackage{draftfigure}
\usepackage{appendix}
\usepackage[colorlinks=true
,urlcolor=blue
,anchorcolor=blue
,citecolor=blue 
,filecolor=blue
,linkcolor=blue
,menucolor=blue
,linktocpage=true
,pdfproducer=medialab
,pdfa=true
]{hyperref}

\def\figureautorefname~#1\null{Fig.\,#1\null}

\def\equationautorefname~#1\null{Eq.\,(#1)\null}

\newcommand{\beq}{\begin{equation}}
\newcommand{\eeq}{\end{equation}}
\newcommand{\bea}{\begin{eqnarray}}
\newcommand{\eea}{\end{eqnarray}}
\newcommand{\ba}{\begin{array}}
\newcommand{\ea}{\end{array}}

\def\m1{M_1}
\def\m2{M_2}
\def\m3{M_3}

\def\ch10{\tilde \chi^0_1}

\def\to{\rightarrow}

\newcommand{\lsim}{\mathrel{\mathop{\kern 0pt \rlap
  {\raise.2ex\hbox{$<$}}}
  \lower.9ex\hbox{\kern-.190em $\sim$}}}
\newcommand{\gsim}{\mathrel{\mathop{\kern 0pt \rlap
  {\raise.2ex\hbox{$>$}}}
  \lower.9ex\hbox{\kern-.190em $\sim$}}}

\def\fbi{\,{\rm fb}^{-1}}
\def\abi{\,{\rm ab}^{-1}}

\newcommand{\ee}{{$e^{-} e^{+}$}}

\newcommand\snowmass{\begin{center}\rule[-0.2in]{\hsize}{0.01in}\\\rule{\hsize}{0.01in}\\
\vskip 0.1in Submitted to the  Proceedings of the US Community Study\\ 
on the Future of Particle Physics (Snowmass 2021)\\ 
\rule{\hsize}{0.01in}\\\rule[+0.2in]{\hsize}{0.01in} \end{center}}

\begin{document}

%\snowmass

\title{Physics at Future Colliders: \\ the Interplay Between Energy and Luminosity}
\author{Zhen Liu}
\email{zliuphys@umn.edu}
\thanks{\scriptsize \!\! \href{https://orcid.org/0000-0002-3143-1976}{0000-0002-3143-1976}}
\affiliation{School of Physics and Astronomy, University of Minnesota, Minneapolis, MN 55455, USA}
\author{Lian-Tao Wang}
\email{liantaow@uchicago.edu}
\affiliation{Department of Physics, University of Chicago, Chicago, IL 60637, USA}
%\date{September 2021}
\begin{abstract}
In this note, as an input to the Snowmass studies, we provide a broad-brush picture of the physics output of future colliders as a function of their center of mass energies and luminosities. Instead of relying on precise projections of physics reaches, which are lacking in many cases, we mainly focused on simple benchmarks of physics yields, such as the number of Higgs boson produced. More detailed considerations for lepton colliders are given since there have been various recent proposals. A brief summary for hadron colliders based on a simple scaling estimate of the physics reaches is also included.

\snowmass
\end{abstract}

\maketitle
\tableofcontents

%\newpage
\section{Introduction}

There have been many proposals of future colliders~\cite{CEPCStudyGroup:2018ghi,FCC:2018evy,Benedikt:2018csr,AlexanderAryshev:2022pkx,deBlas:2018mhx,deBlas:2022aow,Bai:2021rdg,Litvinenko:2022qbd,Litvinenko:2022mrt}. %Dasu:2022nux
As an input to the Snowmass study, we discuss the connection between luminosity and physics reach at future lepton colliders and hadron colliders as a function of the center-of-mass energy. This is a vast topic. In-depth studies, with extensive simulation taking into account accelerator and detector design details, are needed to generate a full-fledged and precise answer. This is beyond the scope of this short note. Moreover, the detailed designs have not been carried out for some of the more recent proposals included in the Snowmass discussion. At the same time, the current proposals (and the corresponding studies) contain specific run energies and luminosities. It would be helpful to have a sketch of the physics reach beyond those benchmark points, which would help consider variations of the existing proposals and new ones. In this note, we focus on giving such an overview without a specific focus on any proposals.

We consider a broad range of potential collision energies, from 80 GeV to 20 TeV for lepton colliders and up to 100 TeV for pp colliders. Each potential future collider will have a broad physics program. To cover all of them, if not impossible, will go much beyond the scope of this note. There has been much discussion about the physics cases of future colliders in recent years. One outcome is the emergence of a relatively small set of significant physics drivers based on essential questions to be pursued in the beyond-LHC era. Very broadly speaking, they can be put into the following three categories: 1) Understanding the property of the Higgs boson to an unprecedented accuracy; 2) Direct search of new physics resonances near and above weak scale. This is partly motivated by the hierarchy problem but covers new physics beyond that; 3) Testing the WIMP Dark matter paradigm. The proposed future colliders have broader physics programs, and each one of them has its own special strengths. At the same time, these three physics drivers are shared by all of these proposals.

Even after narrowing our focus onto these categories and drawing on available studies as much as possible, 
it is impossible to cover the reach in the range of energy and types of machines of interest here. We will content ourselves to offer a high-level view, with ballpark (crude) estimates based on simplifications and approximate extrapolations. We will lay out the detail of our estimates in each physics topic. 

To put the capabilities of future colliders in perspective, we will show a set of targets for physics output, such as the number of Higgs bosons to be produced per year. These are based on the collection of available studies and simple extrapolations and should be thought of as ballpark estimates. Our results are presented so that it is easy to scale to new targets. Many of the projections will be updated due to the Snowmass studies in the near future, and we will update our targets accordingly. We will also overlay these targets with the proposed run plan of the various future colliders. We note that many run plans are still being finalized in the Snowmass processes. For definiteness, we will use the plan presented at a series of Agora meetings \cite{AgoraCircularee,Agoralinear,AgoraMuon,AgoraHadron,AgoraAdvanced} for this version of the white paper. We will update the figures with the finalized numbers. Part of the motivation of this work is our discussion with the Implementation Task Force in the Snowmass studies on presenting the physics yields of various future collider proposals.

\section{Lepton colliders: sub-TeV case}
\label{sec:lowEee}

Leptons colliders below TeV often have their primary physics goals as precision measurements of the properties of the SM particles. An obvious focus is the Higgs boson. Many proposed experimental programs exists in this energy range, such as CEPC~\cite{CEPCStudyGroup:2018ghi}, FCC-ee~\cite{FCC:2018evy}, ILC~\cite{AlexanderAryshev:2022pkx}, CLIC-380~\cite{deBlas:2018mhx}. 
Meanwhile, many new possibilities are also under consideration, such as a 125 GeV Muon collider~\cite{deBlas:2022aow}, $C^3$~\cite{Bai:2021rdg,Dasu:2022nux}, ReliC~\cite{Litvinenko:2022qbd} and CERC~\cite{Litvinenko:2022mrt}. 
%We will update the figures after the details of the run plans are finalized. 
The main functions of this class of machines are 1) Higgs factory; 2) Z factory; 3) WW, threshold and beyond; 4) $t{\bar t}$, threshold and beyond. In this section, we discuss them in turn. 

\begin{figure}[h!]
   \centering
   \includegraphics[height=0.55\textwidth]{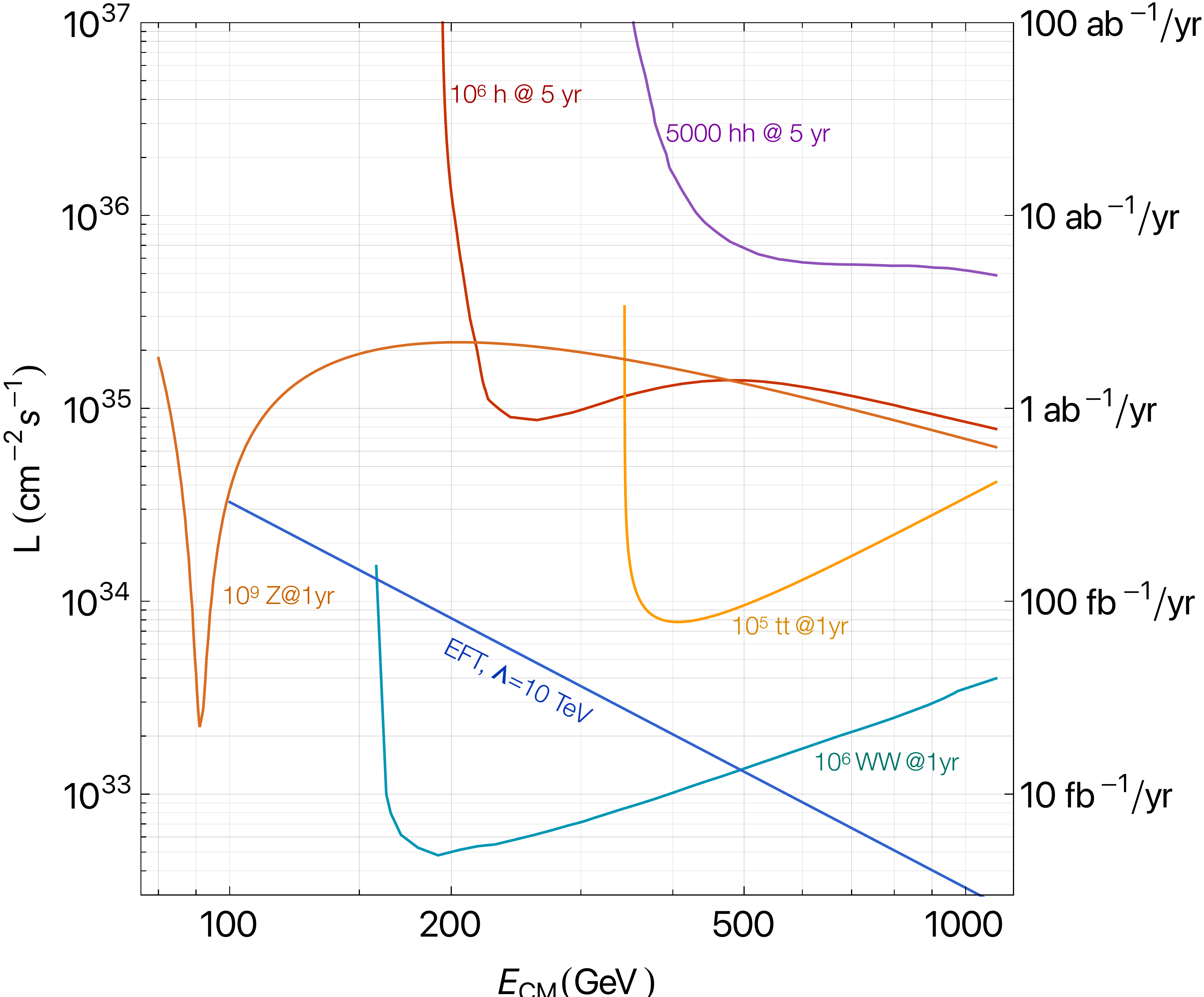} 
       \includegraphics[height=0.55\textwidth]{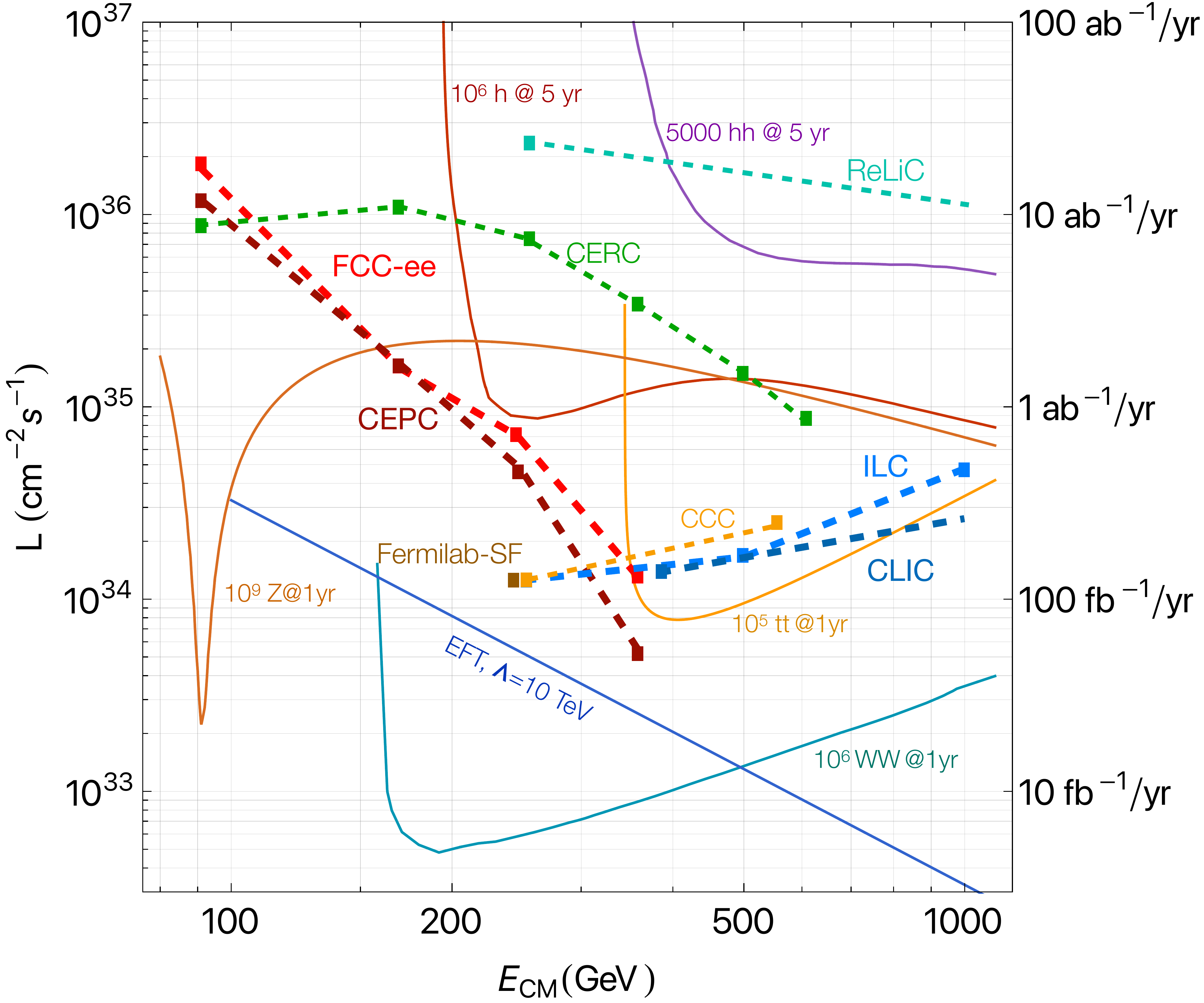} 
   \caption{Top: Required luminosity for lepton colliders for various physics goals in the energy range of 80 GeV to 1 TeV.  %\LTW{put the following in the text, also slightly reworded} 
   The required instantaneous luminosities assume one detector. For experiments with multiple concurrent detectors running, one can multiply its luminosity by the number of detectors to obtain the effective physics yields. 
   %The required luminosities for colliders with more than one detector can be scaled straightforwardly.
   The run plan of the colliders often has various stages to achieve different precision goals. Hence, the luminosities goals displayed here are ``ballpark" figures which loosely correspond to run plans. The actual physics output for each stage of a proposed collider can be worked out from this figure straightforwardly.   
   Bottom: We overlaid the instantaneous luminosities of various proposed lepton colliders in this energy range. The run plans are taken from Snowmass Agora presentations~\cite{Agoralinear,AgoraCircularee,AgoraMuon}.
  % \ZL{Please start the plot of EFT at 100 GeV, instead of 80 GeV. The interference effects around Z-pole is complex and I did not take it into account.}
   }
   \label{fig:lowEee}
\end{figure}
%\clearpage
%\begin{figure}[h!]
%   \centering
%    \includegraphics[height=0.68\textwidth]{Figs/LEeeCollider0321.pdf} 
%   \caption{}
%   }
%   \label{fig:lowEeeCollider}
%\end{figure}

The summary plot is shown in \autoref{fig:lowEee}. We will discuss the corresponding calculations and physics details later in this section. We select six different (minimal) ``target" or ``benchmark" cases for lepton colliders in this energy range to demonstrate the required luminosity. These six target cases are centered around Higgs physics (single Higgs and double Higgs productions), electroweak physics ($WW$, $Z$ and EFT), and top physics ($t\bar t$). Depending on the center-of-mass energy of the lepton colliders, the required luminosities vary mostly in the range  from $10~\fbi$ per year to $10~\abi$ per year. In the bottom panel of \autoref{fig:lowEee}, we overlaid the run plans of many proposals currently under consideration in the Snowmass process. 
%The run plans of many of these proposals are still under development, and updated numbers for Snowmass reports will be finalized in the near future. In this version of our whitepaper, for definiteness, we will take the run plans presented in the Agora meetings\cite{Agoralinear,AgoraCircularee,AgoraMuon,AgoraHadron,AgoraAdvanced}. 
We can see from \autoref{fig:lowEee} that most proposed colliders can cover all these cases reasonably comfortably. Hence, there is room for relaxing luminosities requirements at early stages if constrained by other considerations while still achieving some physics goals. 
%We also show the run plans of different collider proposals in this figure. 

It is important to note the impact of beam polarization here. Beam polarization generally can help reach the same physics goals with lower luminosity. The signal cross-section can increase in the favorable configuration, and the background might also be suppressed. Further, if one can also flip the beam polarization, new physics can be accessed more readily by eliminating various systematic uncertainties through asymmetry observables~\cite{Han:2013mra,Craig:2015wwr,Janot:2015yza}. However, the plot will be jam-packed if we also consider different polarization configurations, defeating the starting point of this study to provide a simple, quantitative, and schematic understanding of the luminosity requirement for future colliders. Hence, we choose to show the cases with {\it unpolarized} beams only. In this regard, the designed luminosity of machines, such as ILC or CLIC, have an effective luminosity higher than plotted here, possibly by a few tens of percent. For instance, the ILC-designed beam polarization of $(+0.8,-0.3)$ will yield 40\% more Higgs bosons than unpolarized beams~\cite{AlexanderAryshev:2022pkx}. 

Next, we discuss each physics goal and the corresponding collider performance in detail.

{\flushleft \bf Single Higgs and Higgs coupling precision.} One of the main goals of low energy lepton colliders is to function as a Higgs factory and measure the Higgs couplings with unprecedented precision. HL-LHC can measure some of the Higgs couplings to an accuracy of a few percent. Hence, a meaningful target for a Higgs factory would be to push the accuracy higher by one order of magnitude to the per mil level. New physics, at scale $\Lambda$, will generically modify the Higgs coupling at the level of 
\bea
\delta = c \frac{v^2}{\Lambda^2}
\label{eq:deviation}
\eea
A per mil level measurement allows us to probe new physics at the level of a few TeV. Numerous detailed studies have supported this generic expectation~\cite{Gu:2017ckc,deBlas:2019rxi}. 

For a specific Higgs coupling, obtaining the projected accuracy at a given collider requires detailed studies by combining all the relevant channels, considering possible degeneracies, and incorporating realistic accelerator and detector conditions. While this has been done for some of the existing proposals~\cite{An:2018dwb,CEPCStudyGroup:2018ghi,AlexanderAryshev:2022pkx,FCC:2018evy,deBlas:2018mhx}, it is not possible to accurately extrapolate to a broader range of energies and luminosities. At the same time, conservatively, $10^6$ Higgs boson would at least be needed even to have a chance of measuring $HZZ$ Higgs coupling to a per mil level. Again, this expectation is borne out by the studies, and current proposals all roughly have $10^6$ Higgs bosons as a target. Given the importance of the Higgs precision program, we anticipate most sub TeV colliders spend at least five years on Higgs runs. Hence, 
we set the first target of one million Higgs bosons in five years. 

We show in \autoref{fig:lowEee} the required luminosity for single Higgs production  (red line). Typical $e^+e^-$ machines access such physics through the $ZH$ associated production, reaching a minimal luminosity requirement with the center-of-mass energy around 240-250~GeV. Beyond $ZH$ associated production, the $WW$-fusion starts to dominate the production at around $E_{\rm CM}=400$~GeV. Its importance is already revealed even at somewhat lower energies where it is not the dominant production process. For instance, some circular lepton colliders such as the FCC-ee and the CEPC plan to run at around 360~GeV center-of-mass energy. Running at this energy, the $WW$-fusion production of Higgs provides complementary information about Higgs couplings and help reduce the correlations between measurement relying on $ZH$ production. Linear colliders can access the Higgs physics at higher center-of-mass energies. For example, the CLIC would start at $E_{\rm CM}=380$~GeV. The $WW$-fusion and $ZH$ associated production would both play essential roles. The detailed comparison of the Higgs precision physics reach can be found in various studies, e.g., Ref.~\cite{deBlas:2019rxi}. There is also a proposal for a 125 GeV muon collider Higgs factory. At this collider, the production mode for Higgs differs from this plot, and hence we did not show it here. For $10^6$ Higgs within 5 years, one needs about $14\fbi/{\rm yr}$ ($1.4\times 10^{32}~{\rm cm}^{-2}{\rm s}^{-1}$) with a resonance scan strategy developed in Ref.~\cite{deBlas:2022aow}. Beyond Higgs coupling precision, the lepton collider Higgs program also provides intriguing opportunities in probing Higgs exotic decays~\cite{Liu:2016zki,Carena:2022yvx}, complementary to the hadron collider program~\cite{Curtin:2013fra,Cepeda:2021rql}.

{\flushleft \bf Double Higgs and Higgs self-coupling.} 
Another important measurement is the Higgs self-coupling. We distinguish it from the discussion above since it requires good statistics in the double-Higgs final state. The extraction of the Higgs self-coupling requires consistent treatment of all relevant couplings that could affect the $HH$ process at the lepton collider, as many single Higgs couplings can also affect the rates. Many studies have shown that a TeV lepton collider could extract Higgs self-coupling at around 10\% level. 
Beyond the double Higgs production, lepton colliders could also access Higgs self-coupling through loop-induced processes. Ref.~\cite{DiVita:2017vrr} shows that in a consistent SMEFT framework, one can extract the Higgs self-coupling to around 40\% without the $HH$ processes at low energy Higgs factories. 
%Theprecision improves to around 10\% with TeV runs \LTW{we mean improving to 10 percent with the HH processes, right?}. 
To set a target for the $HH$ process, we show the required luminosity for $5\times 10^3$ $HH$ within 5~years. This calculation included the $ZHH$ associated production, VBF $HH$ production, and $t\bar t HH$ productions. The former two processes are dominant at an energy below and above around 800~GeV, respectively.

{\flushleft \bf Precision $Z$. }
Circular \ee colliders offer the possibilities of a high statistics $Z$ factory. Such a $Z$ factory also has a broad physics program. First of all, as demonstrated by LEP-I and SLC, a $Z$-pole program brings in a suite of electroweak precision observables. LEP-I produced about $10^7$ $Z$ bosons. To be significantly better, a new $Z$-factory would need to produce at least $10^9$ $Z$ bosons, dubbed ``Giga Z".

The precision electroweak program is an essential input for other physics quantity extractions, such as the Higgs and top quark couplings. Studies~\cite{Fan:2014vta,DeBlas:2019qco} show that a ``Giga Z" provides sufficient precision to not significantly hinder the physics extraction of other physics programs, making it a good benchmark. A further improvement would require significantly better measurements of other input parameters of the electroweak precision fit, motivating runs at the $WW$ and $t \bar t$ thresholds (discussed later), and increasing the Z-pole statistics.

The $Z$-pole physics as precision electroweak physics can also be accessed in other processes, though less efficiently. For instance, through the ``radiative return" process, one can also measure the $Z$-boson couplings to various final states, and through the EFT contact operators, one can probe four-fermions interactions effectively. In \autoref{fig:lowEee}, we show the luminosity requirement to produce $10^9$ (nearly) on-shell $Z$ boson around $Z$-pole and through the ``radiative return" process. We also include other contributions of diboson processes, but for the region of interest, they are never dominant. The radiative return calculation is based upon the framework and beam structure-function specified in Ref.~\cite{Greco:2016izi,Jadach:2015cwa,Jadach:2000ir}.

In addition to the electroweak precision measurements, a Z factory is both a $\tau$ factory and a $b$ factory. Current proposals of Z factories, such as FCC-ee and CEPC, often set $10^{12}$ Z as a target. This can further improve the sensitivity to rare decays and the potential for them as $\tau$ and $b$ factories \cite{CEPCStudyGroup:2018ghi,FCC:2018evy}, which can be competitive to other $\tau$-charm factories~\cite{Shi:2020nrf} and b factories~\cite{Belle-II:2018jsg} running at much lower energies. 
Moreover, it produces $\tau$ and $b$ at a much more significant boost, which can be beneficial in specific channels. 
Rare Z decay also offers a great opportunity in probing a broad range of dark sector models~\cite{Liu:2017zdh}. The reach-scale linearly with the total number of $Z$s for distinct final states that have a negligible background. 

{\flushleft \bf Precision $WW$ threshold physics. }
$WW$ production near the threshold is another important physics channel at lepton colliders for electroweak precision physics. 
$W$ mass measurement is crucial for interpreting the electroweak precision observable on the $Z$-pole. The precision of the W mass measurement by colliders has an error bar of $\delta m_W \simeq 12 $ MeV \cite{ParticleDataGroup:2020ssz}. The recent result reported by the CDF collaboration improve the precision to $\delta m_W =9.4$ MeV~\cite{CDF:2022hxs} (with an intriguing deviation from the SM electroweak fit and other collider measurements~\cite{ParticleDataGroup:2020ssz,deBlas:2022hdk,Gu:2022htv,Fan:2022yly,Bagnaschi:2022whn}). The HL-LHC is expected to improve this further to about $\delta m_W \simeq 8 $ MeV by the end of HL-LHC \cite{Baak:2014ora,Awramik:2003rn}.  A $WW$ threshold (with precision better than %$\delta m_W \simeq$ \ZL{\st{1} 3} 
a few MeV 
%\ZL{--I couldn't find a 1 MeV number from references and I've heard about 3 MeV. ILC had 5 MeV projections in the old days.} \LTW{Current CEPC and FCC-ee both target about 1 MeV. CEPC CDR is about 3 MeV. I change it to a couple of MeV just to be not too precise, which is consistent with the qualitative nature of our discussion}
) scan would be a decisive W mass measurement. It is also indispensable for realizing the full potential of electroweak precision measurement at a Tera-$Z$ factory~\cite{Fan:2014vta,CEPCStudyGroup:2018ghi,FCC:2018evy}. %The precision $W$-mass measurement can be uniquely accessed via the threshold scan.

At the same time, many new physics generate deviations in the $W$ boson couplings, such as the TGC. This can leads to signals in the $WW$ final state. Since the effect of such new physics tends to grow with energy, it would be beneficial to have good statistics in the WW final states at higher energies. 
Based on these, about $10^6$ $WW$s would be a good target for running close to the threshold. In \autoref{fig:lowEee}, we show the luminosity requirement for $10^6$ $WW$s for one year of running. One can also achieve precision measurements, e.g., on TGC, via $WW$ pairs beyond threshold scan. In this sense, the $WW$ measurement for EFT can be simultaneously realized in the Higgs factory and higher energy runs.

{\flushleft \bf Precision Top quark physics.} 
The motivation to have a good scan of the $t {\bar t}$ threshold is similar to that for the $WW$ threshold since top mass is a crucial input to the electroweak precision fit. Hadronic uncertainties will limit the top quark mass measurement at hadron colliders to $\delta m_t \sim 10^2$~MeV. Threshold scan with order $10^5$ $t \bar t$ pairs can push the precision on the top mass down to around 10~MeV~\cite{Simon:2019axh}.
Many new physics models also predict deviations in top quark gauge couplings. Due to large backgrounds and theoretical uncertainties, the HL-LHC will have limited sensitivities to top quark gauge couplings. The lepton colliders could probe the top gauge couplings and top EFT operators to a good precision through direct pair production and their angular correlations. To achieve 10~MeV top mass precision and sub-percent level top quark coupling precision, we set a target of $10^5$ $t\bar t$ per year, shown in the orange curve in \autoref{fig:lowEee}. The $s$-channel Drell-Yan-like processes drive the production of low-energy lepton colliders. When we move to high-energy lepton colliders, the production will be driven more by the VBF process~\cite{Han:2020uid,Costantini:2020stv}.

{\flushleft \bf EFT.} Lepton colliders offer a clean environment for the search for new physics. At the same time, the direct production of the new physics particles is limited by the center-of-mass energy. Hence, the direct production of high-scale physics is unlikely to be a significant physics driver in the range of energies considered here. 

On the other hand, integrating-out new physics leads to higher dimensional EFT operators. Such EFT operators will modify the SM couplings on the order shown in \autoref{eq:deviation}. Hence, precision measurement of Higgs, $Z$, $W$, and top couplings discussed above probe such new physics. At the same time, in the appropriate channels, the new physics effect can grow with energy as $(E/\Lambda)^2$. Hence, it can be beneficial to measure them at high energies. The EFT probe is different in its sensitivity scaling with collider energy compared to the on-shell precision measurement. A typical example of such new physics is a heavy $Z'$ resonance. Integrating out such a resonance will generate a dimension-6 four-Fermi operator, which in term modifies the high energy distribution of $\ell^+ \ell^- \to f \bar f$ \footnote{Heavy vector resonances can also have other final states such as $WW$ and $Zh$. The scaling of the reach in these channels would be similar.}. 
For valid momentum expansion of the EFT, at linear order, for a vector-vector four-fermion $e^+e^-f\bar f$ operator ($\bar e \gamma_\mu e \bar f \gamma^\mu f$) with coefficient $c_{\rm EFT}$ and scale $\Lambda^2$,
\beq
\Delta \sigma(e^+e^-\to f \bar f) \propto  \frac {c_{\rm EFT}}{\Lambda^2}, 
\eeq
where $c_{\rm EFT}$ parameterizes the coupling between the new physics and the SM fermions.
Meanwhile, the SM background scales as
\beq
\sigma(e^+e^-\to f\bar f) \propto \frac 1 {E^2}.
\eeq
Hence, for the fixed target scale of EFT sensitivity, assuming dominance of statistical uncertainty, the required luminosity scales as $1/E^2$. In \autoref{fig:lowEee} we set a EFT scale of $\Lambda/\sqrt{c_{\rm EFT}}=10$~TeV to demonstrate the required luminosity generic $10$-TeV scale generic new physics in EFT. Specifically in \autoref{fig:lowEee} and \autoref{fig:highEee}, we choose $f$ to be bottom quark.

\section{Lepton collider: TeV and Beyond}
\label{sec:highEee}

Beyond sub-TeV lepton colliders, the proposals of realizing multi-TeV, even multi-tens of TeV lepton colliders through various techniques show great promise in exploring high energy particle physics. These exciting possibilities include CLIC~\cite{deBlas:2018mhx} (and envisioned CLIC-like upgrade possibilities for lepton colliders such as ILC and $C^3$~\cite{Dasu:2022nux}), muon collider~\cite{Aime:2022flm}, plasma weak field acceleration~\cite{Fuchs:2022izw}, etc. In this section, we focus on high energy lepton colliders with $E_{\rm CM}$ in the range of 1 - 20 TeV\footnote{The upper limit of the energy is mainly set by what's being proposed during the Snowmass process. The lesson we learned here can also be applied to higher energy lepton colliders, with the targets extrapolated in a straightforward manner. }. Here, the primary goal would be searching for heavy new physics resonances. At the same time, high energy lepton collider can still contribute to the measurement of the Higgs coupling, such as Higgs precision coupling measurements, top Yukawa coupling, and Higgs self-coupling. 
Since \ee and $\mu^- \mu^+$ colliders have very similar reaches in this range of energies,  we do not distinguish them. 
%\ZL{I don't know whether we should comment on Photon collider here or not. The next sentence is your writing.} Photon collider, on the other hand, has a somewhat narrower physics program since it can only produce charged particles. Besides this limitation, they will be similar to the lepton colliders. 

\begin{figure}[h!]
   \centering
     \includegraphics[height=0.52\textwidth]{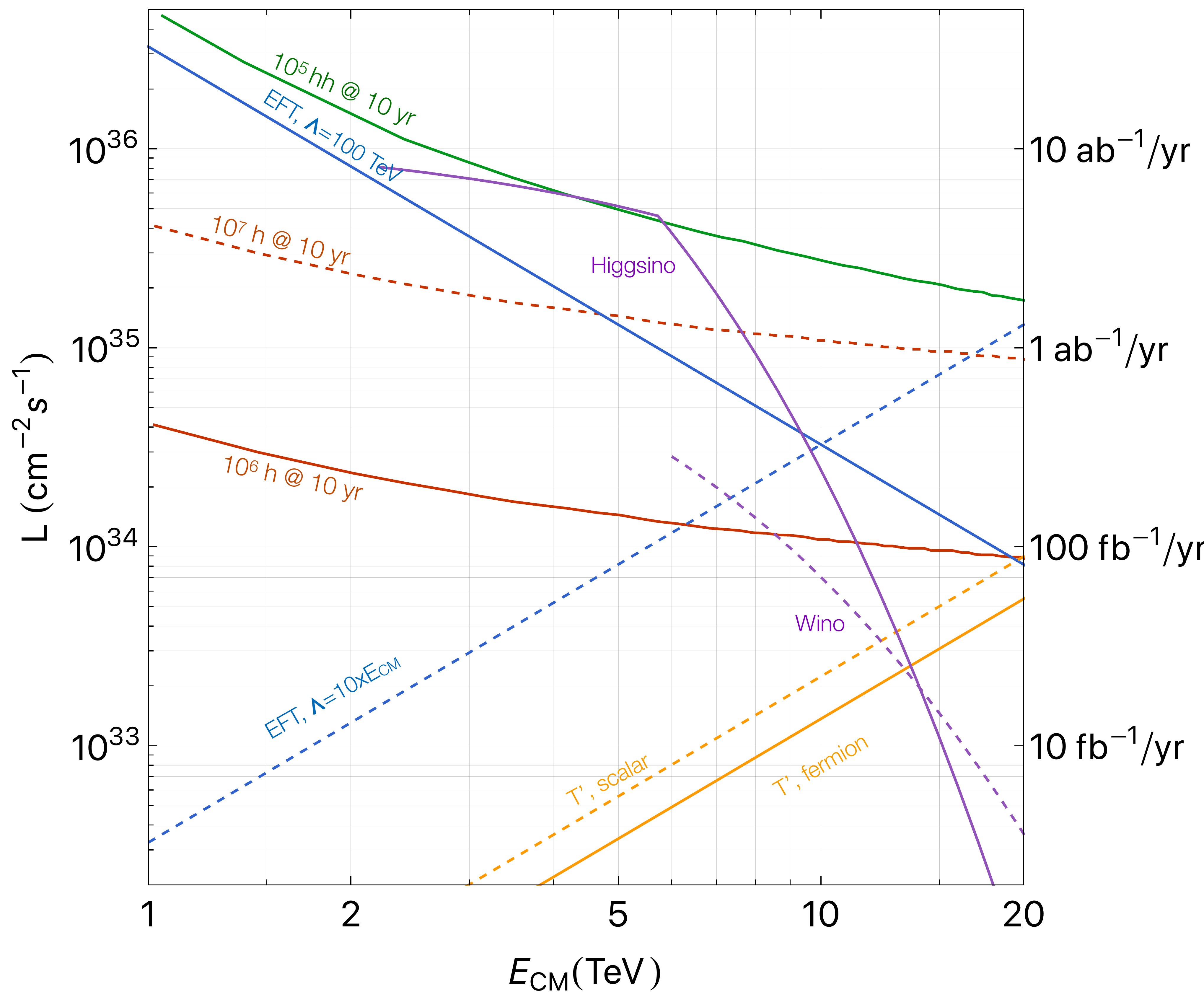} 
    \includegraphics[height=0.52\textwidth]{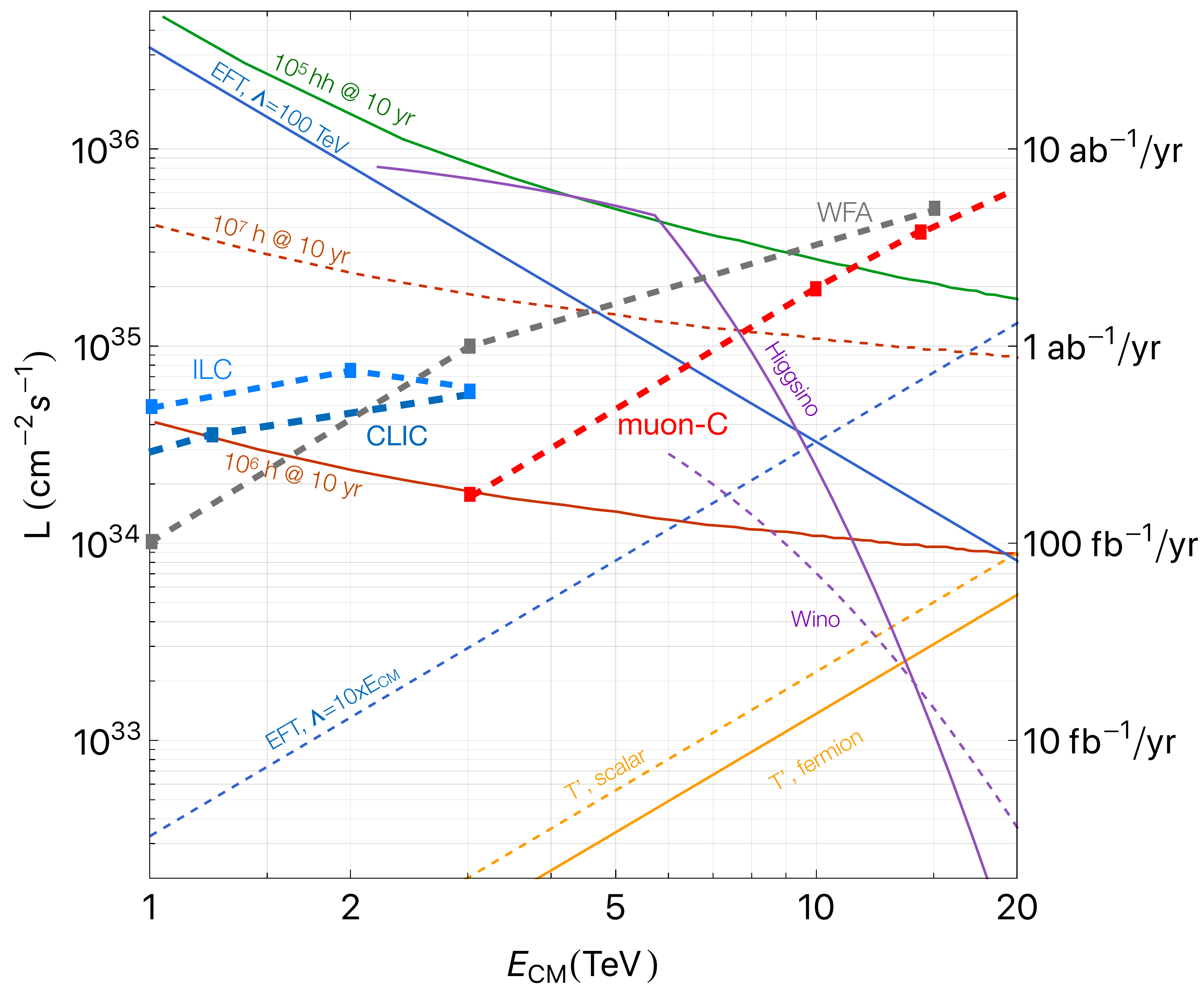} 
   \caption{Required luminosity for lepton colliders for various physics goals in the energy range 1 TeV to 20 TeV, assuming one detector for each experiment. 
   %~\ZL{Why did you stop plotting at 20~TeV? Do you want to to change the discussion slightly from 30 TeV to 20 TeV in the caption and text? I saw your footnote2 but muon collider } 
   For high-energy lepton colliders, while it is possible to have a staged approach to run at different energies, such plannings are still at very early stages. 
   %one usually has simultaneous access to all physics goals. 
   Hence we assumed a uniform run time of 10 yrs. The luminosity requirement for 5$\sigma$ discovery of the benchmark DM scenarios Higgsino and Wino are shown in purple lines. The disappearing track search dominates the Dark Matter discovery luminosity curves. The exception is for Higgsino with lepton collider energy below 6~TeV, where the boost of the Higgsino is insufficient to generate the signal, and the missing energy search dominates the search; for details, see Refs.~\cite{Han:2020uak,Han:2022ubw}. Bottom: We overlaid the instantaneous luminosities of various proposed lepton colliders in this energy range. The run plans are again taken from Snowmass Agora presentations \cite{Agoralinear,AgoraCircularee,AgoraMuon,AgoraAdvanced}.
   }
   \label{fig:highEee}
\end{figure}

The summary of our results are shown in \autoref{fig:highEee}. In the following, we will  discuss the additional physics cases and considerations beyond those of the lower energy lepton colliders discussed in the previous section~\ref{sec:lowEee}. 

{\flushleft \bf Single and Double Higgs.}  The Higgs boson precision program is an essential component of a high-energy lepton collider as well. The high center-of-mass energy allows various new opportunities around precision Higgs physics. First, one can produce more Higgs bosons in the VBF dominant processes at high energy lepton colliders. Hence, we show two benchmark targets of $10^6$ and $10^7$ Higgs bosons, shown in solid red and dashed red lines in \autoref{fig:highEee}. We include various single and double Higgs production modes to determine such target luminosity curves. However, the result is unchanged if we only include VBF WW-fusion Higgs production in the region of interest. From the lineshape, we can see that as one goes to higher center-of-mass energy, the required luminosity is lower due to the logarithmic enhancement in such processes. One should note that the Higgs boson at high energy lepton collider might be associated with a more significant boost, and the combinatorics background can be larger. %\LTW{lower?} \ZL{larger as there is combinatorics. E.g., when searching for VBF-Higgs-4j, there will be various hadronic diboson backgrounds. Or when one searches for diHiggs, hadronic multiboson processes will definitely kick in.} than at the lower energy colliders.

The large center-of-mass energy of high energy lepton colliders also enables copious production of the double Higgs processes, even triple Higgs~\cite{Han:2020pif,Chiesa:2020awd}. Measuring these processes enables us to determine the Higgs potential further. Here in \autoref{fig:highEee} we set a target of $10^5$ double Higgs process, which should be sufficient for one to extract Higgs self-coupling at the percent level. Again, the double Higgs process is dominated by the VBF process, and we can see the required luminosity is high.

{\flushleft \bf EFT.} 
Compared with the lower energy lepton colliders, running at higher energies can extend the reach to new physics encoded in SMEFT  even further by measuring its contribution to high energy distributions. 
As argued in the previous section, for a typical four-fermi contact interaction, the required luminosity for a fixed target scale of $\Lambda/c_{\rm EFT}$ scales as $1/E^2$. For high-energy lepton colliders, our target scale would be more prominent. In \autoref{fig:highEee}, we set two target scales; one is fixed at 100~TeV, and the other is a factor of 10 higher than the center-of-mass energy. We show the required luminosities in solid and dashed blue curves in these two setups.

{\flushleft \bf Top partners, $T'$. }
The hierarchy problem is undoubtedly one of the most prominent  physics motivations for new physics beyond the Standard Model, and addressing it is a leading physics driver for future colliders. Among the new physics particle associated with the hierarchy problem, the top partner is probably the most important one due to the significant role the top quark played in the dynamics of the electroweak symmetry breaking. It is one of the leading physics targets for the upcoming runs of the LHC and will continue to be so for future colliders in the coming decades. Hence, we will use the top partner as a benchmark to characterize the capability of high-energy lepton colliders in new physics searches.

High energy lepton colliders are optimal in searching for top partners. In most models of top partners, it has either the same or similar quantum numbers as the SM top quarks. Hence, it can be produced copiously at the lepton colliders through their EW gauge interactions in the Drell-Yan-like process. Since the decay of the top partners typically gives energetic and visible final states, we expect they can be identified and separated from the background in the lepton collider environment. With small statistics of 20 signal events, one should be able to discover them. We expect the reach of the top partner at the lepton collider should be close to the kinematic threshold. With this in mind, we set a target top partner mass to be at 90\% of the pair production kinematic threshold, $2 m_{T'} = 0.9 \times E_{\rm CM}$,  and show the required luminosity for a generic scalar or fermionic top partners with dashed and solid yellow curves in \autoref{fig:highEee}. Any top partner with a mass less than this target should be discoverable for the given luminosity. The production rate will be slightly higher with the removal of threshold suppression. In principle, a top partner with a mass at the weak scale could have a sizable background from the SM processes, depending on its primary decay channel. However, this possibility is highly disfavored with the current LHC search results. 

{\flushleft \bf WIMP Dark matter.} 
Testing the WIMP (Weakly Interacting Massive Particle) paradigm of dark matter is another main physics driver for future colliders. Among various possible candidates, the minimal model would be dark matter as a member of an electroweak multiplet. Such dark matter multiplet can be produced copiously at the high energy lepton colliders. However, unlike in the case of top partners, the signal is more difficult to detect. One has to rely on relatively soft objects, and additional radiated SM particles in the event to identify the signal. The background would be significant at lepton colliders. It is not possible to determine the reach from simple estimates. Fortunately, there have been relatively detailed studies in this case~\cite{Han:2020uak,Han:2022ubw,Capdevilla:2021fmj,Bottaro:2021snn,Bottaro:2021srh}. The signals can be categorized into general and inclusive missing energy searches. One can look for the mono-photon, mono-$Z$, mono-$W$, and even mono-lepton at lepton colliders to hunt for WIMP dark matter. The Drell-Yan-like pair productions dominate the WIMP dark matter production. However, for WIMP dark matter mass somewhat smaller than the kinematic threshold, the VBF process can dominate. The VBF process produces different kinematics and allows for interesting and clean signals such as mono-lepton or VBF di-leptons~\cite{Han:2020uak,Han:2022ubw}. 

The fermionic electroweak doublet and triplet are particularly important among the WIMP candidates. They are also the primary DM candidates in the Minimal Supersymmetric Standard Model, known as pure-Higgsino and pure-Wino, respectively. Beyond the inclusive missing energy searches, one can rely on the somewhat long-live charged state in the multiplet. A typical lifetime in the rest frame of the pure Higgsino and pure Wino is around $0.02$~ns and $0.2$~ns and thermal dark matter target mass of 1.1~TeV and 2.8~TeV, respectively. One can search for the corresponding ``disappearing track" signature. The availability of this signal is somewhat model dependent as it is very sensitive to the mass splitting with the electroweak multiplet. However, if available, the disappearing track empowers a much higher signal to background ratio and hence a more effective search. 

We show the required luminosity as a function of the lepton collider center-of-mass energy for Higgsino and Wino in solid and dashed purple lines, respectively. At low center-of-mass energy, the search sensitivity is driven by the inclusive missing energy searches, as the boost of the pair-produced charged Higgsino is low. Hence, the lifetime is not long enough to support a high efficiency of the ``disappearing track" signatures. Beyond 6~TeV center-of-mass energy, the search sensitivities are driven by the disappearing track searches. Different lepton colliders will have different sources of background for such a signature, and future studies will be of particular importance with concrete beam and detector designs. We can see from the figure that a 10~TeV lepton collider could achieve the Higgsino and Wino dark matter goal robustly.

\section{High energy proton proton collider}

\begin{figure}[h!]
    \includegraphics[width=0.8\textwidth]{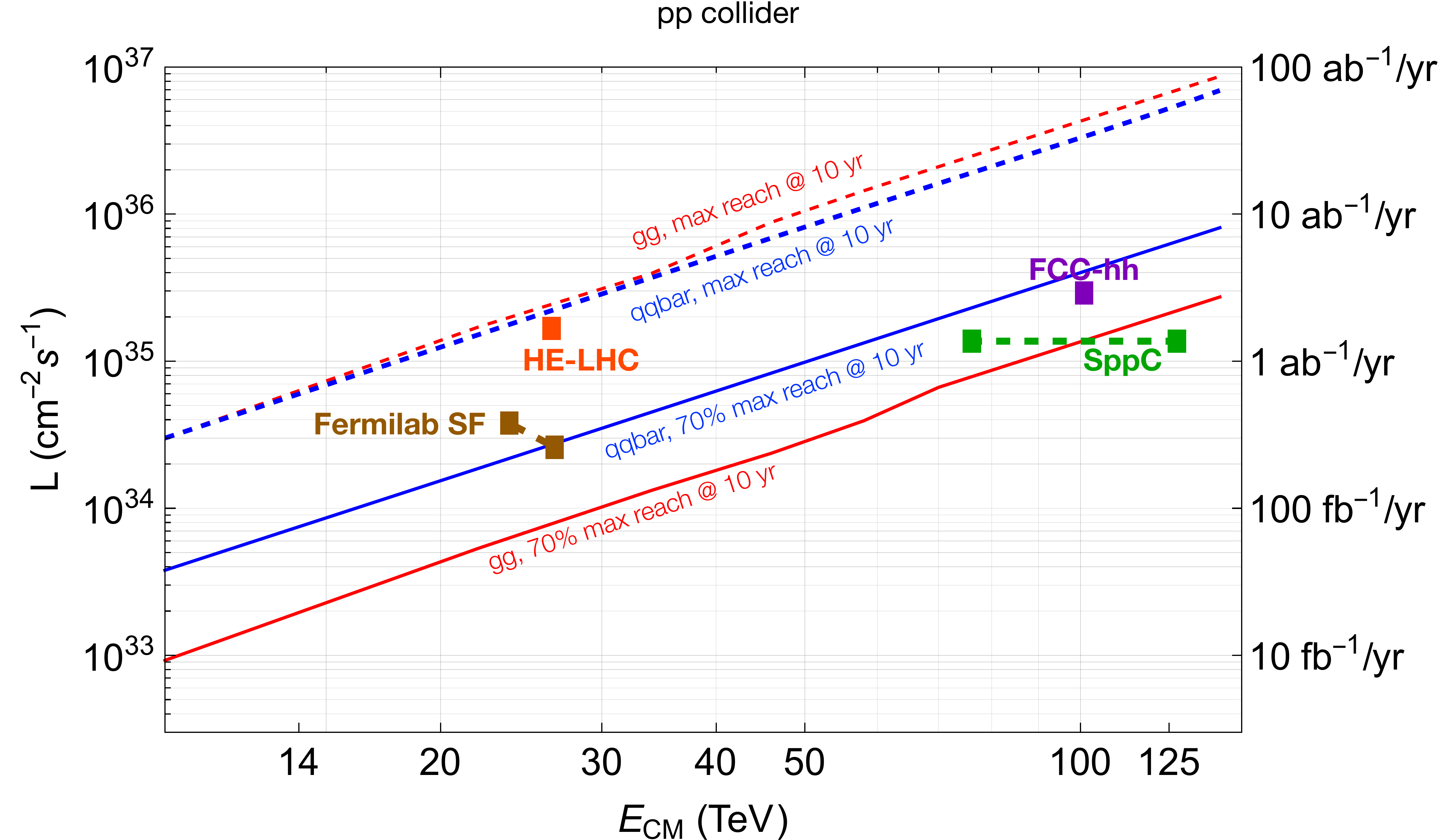}
    \caption{Luminosity requirements for high energy hadron colliders, for possible initial states (gg, red; $q \bar q$ blue). One set of goals, the optimal case, is to have the ultimate reach scale linearly with the center-of-mass energy. This would require the luminosity scale as $E_{\rm CM}^2$, similar to the lepton collider. At the same time, to get most of the increase in the reach, say $70 \%$ of the optimal reach, the required luminosity is much lower. This is a consequence of the scaling of parton luminosity as a function of the parton center-of-mass energy. The scaling of the reach shown here is done by statistics and using parton luminosity~\cite{parton_scaling}. We used the reach at the HL-LHC as a reference point for the extrapolation. For the gluon-gluon initial state dominated process, we assumed the reach of the mass of new physics at the HL-LHC of 3 TeV (approximately 1.5 TeV for pair production). This could be similar to the case, for example, the stop. For $q \bar{q}$ initiated processes, we assumed a reach at the HL-LHC of 1 TeV (which would be about 500 GeV for pair production). This would be similar to the electroweak states. Changing the assumption of HL-LHC reach will give rise to some differences but affect the qualitative feature. The luminosities of several proposed future hadron colliders, taken from Agora presentations \cite{AgoraHadron}, are shown. 
    }
    \label{fig:ppLumiE}
\end{figure}

High energy proton-proton colliders~\cite{Benedikt:2018csr,AgoraHadron}, with $E_{\rm cm}$ up to the range of 100 TeV, is a promising way of exploring the energy frontier. Here, the physics targets for specific proposals have been carefully studied \cite{CEPC-SPPCStudyGroup:2015csa,Benedikt:2018csr}, including the luminosities needed to achieve them \cite{Hinchliffe:2015qma}. In this section, we focus on the interplay between luminosity and energy for a broader range of energies. 

It is more challenging to make projections for the hadron colliders, especially for searches dominated by systematics. In this case, the prospects depend sensitively on the assumption of potential improvement in these systematics. Some of these can undoubtedly be improved with the accumulation of data, such as those channels dominated by the statistics of a sideband. At the same time, a lot could depend on the details of the accelerator and detector. 

At the same time, it is possible to make some rough estimates for searches based on the behavior of parton luminosity and statistics \cite{parton_scaling}. 
This method, and a discussion of the qualitative behavior of the reach, are summarized in Appendix~\ref{app:had_scaling}. By the nature of this estimate, we will not be able to take into account subtle kinematic effects. The main factor which determines the reach is the composition of the initial state that dominates the production. For new physics charged under color, typically, the gluon-gluon initial state would give the dominant contribution. For example, this is the case for the stop in supersymmetry\footnote{A possible exception is diquark production, to which valance quark gives the dominant contribution.}. At the same time, $q \bar{q}$ initial state dominants the contribution to new physics, which are neutral under color, both for the case of single production (such as $Z'$) or pair production (such as electroweak multiplets).

\begin{figure}[h!]
    \includegraphics[width=0.45\textwidth]{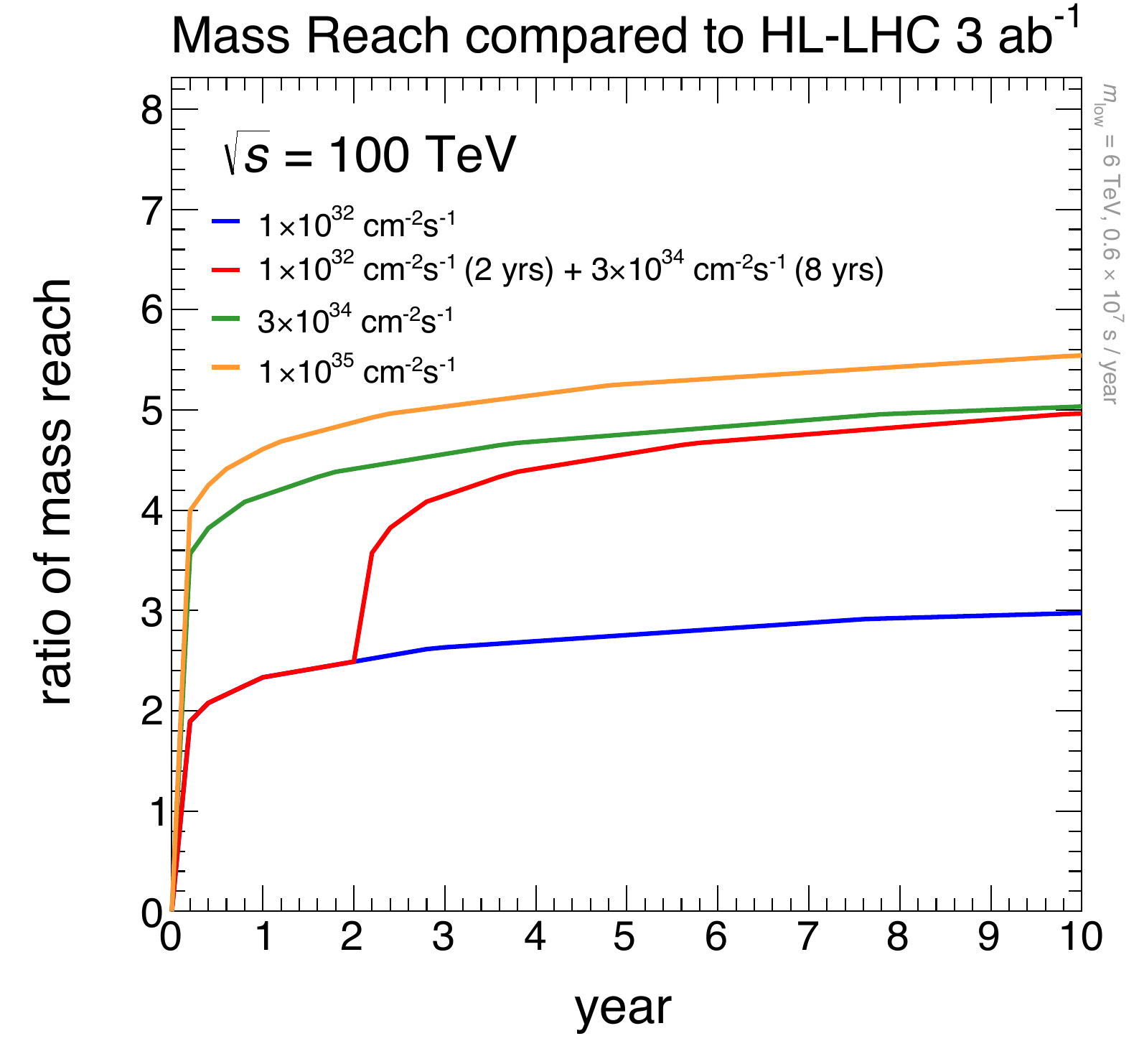}
        \includegraphics[width=0.40\textwidth]{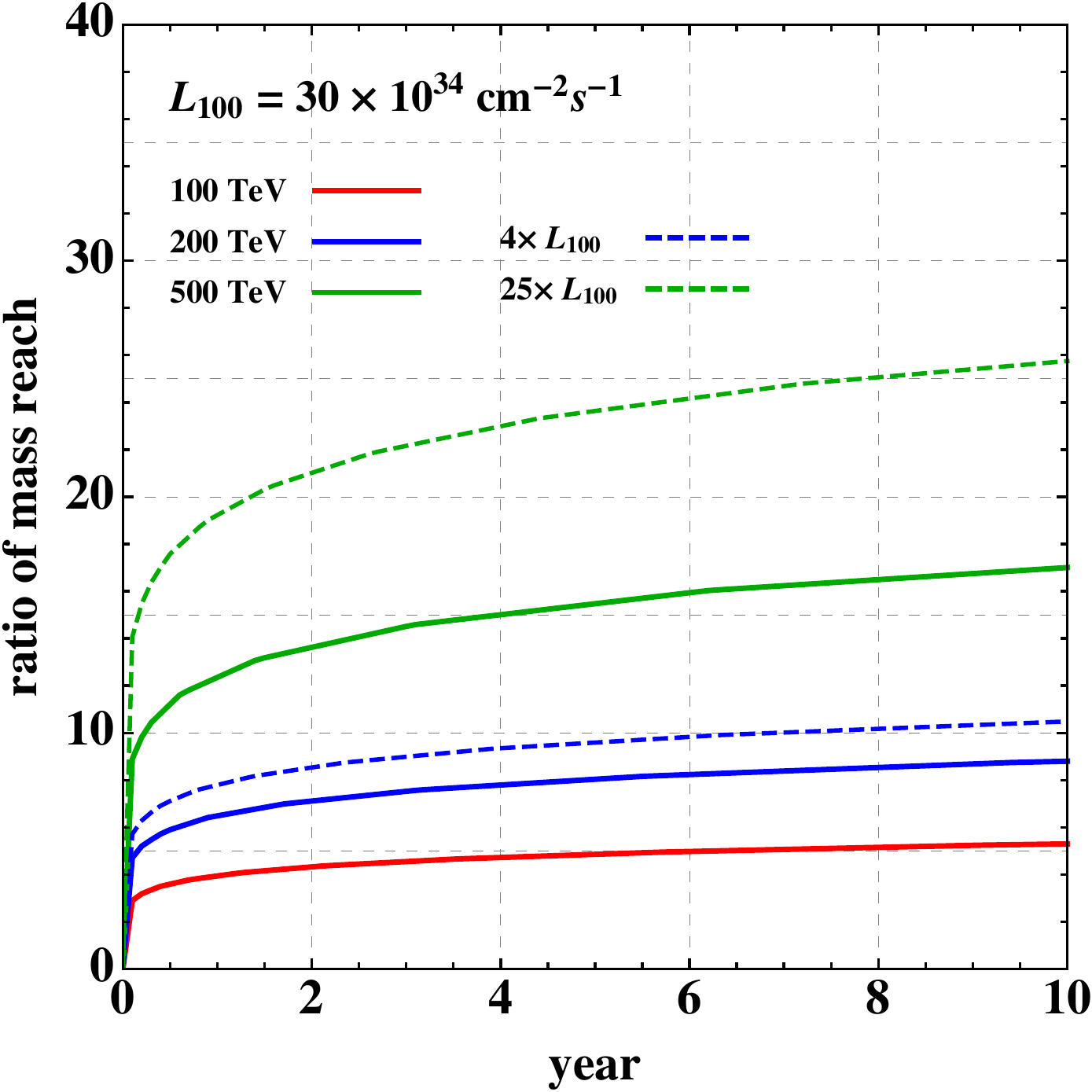}
    \caption{Mass reach at high energy hadron colliders in comparison with the reach at HL-LHC \cite{Hinchliffe:2015qma}. The case of gg initial state is used as an example. During the initial stage of the run, there is a rapid gain of reach for a relatively small amount of integrated luminosity.  
    }
    \label{fig:ppLumiTime}
\end{figure}

Our result is presented in \autoref{fig:ppLumiE}. In the following, we make a couple of observations.
\begin{itemize}
    \item Given two colliders with the center of mass energies $E_1$ and $E_2$, the corresponding reach of the masses of a particular new physics particle are denoted as $M_1$ and $M_2$, respectively. Ideally (the optimal case), one would like to have the reach of new physics scale linear with the center-of-mass energy, $M_2 /M_1 = E_2 / E_1$. With a mild assumption on the production rate scaling, it is straightforward to see we need ${\cal L}_2 / {\cal L}_1 = E^{2}_2 / E^2_1$ (see Appendix~\ref{app:had_scaling}), which  can be seen in the dashed curves in \autoref{fig:ppLumiE}. This could be a large step in luminosity increase for a large increase of the center-of-mass energy.  
    \item Due to the nature of the parton luminosity as a function of parton center-of-mass energy, there is a rapid gain in reach for a relatively small amount of luminosity. This can also be seen in the sharp rises during the early stages of the run, shown in \autoref{fig:ppLumiTime}. Hence, to achieve a somewhat lower goal for the mass reach enhancement, for example, $70 \%$ of $E_2/E_1$, one needs a significantly smaller amount of data, shown as the solid lines in \autoref{fig:ppLumiE}. 
    \item The parton luminosity for the gluon-gluon initial state falls as a function of the parton center-of-mass energy faster than the $q \bar{q}$ initial state. As a result, the luminosity needed for obtaining a significant fraction of the maximal reach in the gluon-gluon initial state-dominated processes is less than that of processes dominated by the $q\bar{q}$ processes (see Appendix~\ref{app:had_scaling}). This is also shown in \autoref{fig:ppLumiE} as the difference between the solid red and blue lines.
\end{itemize}

\section{Summary}
In this white paper, we provide a broad-brush picture of the physics output of future colliders as a function of their center of mass energies and luminosities. The key results for colliders at different energies can be found in \autoref{fig:lowEee}, \autoref{fig:highEee}, and \autoref{fig:ppLumiE}. This is an input to Snowmass studies, in particular for the Implementation Task Force, which helps present the physics yields of various future collider proposals.

\acknowledgements{The authors thank Xing Wang for helpful discussion and Matthew Low for help in preparing the results for hadron colliders. We would like to thank the members of the Implementation Task Force for their discussions and suggestions. ZL is supported by the U.S. Department of Energy (DOE) under grant No. DE-SC0022345. LTW is supported by the DOE grant DE-SC0013642.
ZL acknowledges the support of the Munich Institute for Astro-, Particle and BioPhysics (MIAPbP), where the final stage of this work is conducted, which is funded by the Deutsche Forschungsgemeinschaft (DFG, German Research Foundation) under Germany´s Excellence Strategy-EXC-2094-390783311.
}

%\newpage
\appendix
\section{Scaling for the reach at hadron colliders}
\label{app:had_scaling}

The basic feature of luminosity needed as a function of $E_{\rm CM}$ can be understood as follows. Consider two colliders with center-of-mass energies $E_1$ and $E_2$, with integrated luminosities ${\cal L}_1$ and ${\cal L}_2$, respectively. The reach in certain channel are $M_1$ and $M_2$ for the two colliders. Since the parton luminosities scale as a power law of $\tau = \hat s /E^2_{\rm CM}$, under the assumption that one needs the same number of signal events to obtain a limit or see an excess, we have
\footnote{Here we assume the partonic matrix element squared is independent of $\sqrt{\hat s}$, which is true for a broad class of processes, e.g., typical $2\rightarrow 2$ process through marginal operators, resonance productions with fixed couplings (total width of the resonance proportional to mass). }
\bea
\frac{1}{M_1^2}\frac{1}{\tau_1^a} {\cal L}_1 \simeq \frac{1}{M_2^2} \frac{1}{\tau_2^a} {\cal L}_2,
\eea
where $\tau_{1,2} = M^2_{1,2}/E^2_{1,2}$.
\footnote{Here we assume the same scale of the signal rate and background rate, as a function of partonic center of mass energy $\sqrt{\hat s}\sim M$.}
If we take HL-LHC as a base point for extrapolation, $E_1=14 $ TeV and ${\cal L}_1 = 3 $ ab$^{-1}$. $M_1$ would be the maximal reach achievable at the HL-LHC. $a$ is typically in the range of $3\sim5$.  With this, we obtain
\bea
\left(\frac{\tau_2}{\tau_1}\right)^{2+2a} \simeq \frac{E_1^2}{E_2^2} \frac{{\cal L}_2}{
{\cal L}_1}. 
\eea
If we want to achieve the maximal reach, i.e., the reach scale with the center-of-mass energy, then $\tau_2 = \tau_1$. In this case, the luminosity needed is proportional to the $E_{\rm CM}^2$, as expected. On the other hand, if we would like only to achieve a fraction $x$ (say $x=70 \% $) of the maximal reach, we need a factor of $x^{2+2a}$ less luminosity. This is significantly less if $a$ is large. In this case, the process with initial state parton luminosity with a smaller $a$ would need more luminosity, as in the case of the $q \bar{q}$ in comparison with $gg$ initial states. Of course, this approximation is crude and ignores a couple of effects. First of all, naively, $\tau_1$ would cancel in this extrapolation, as the result will only depend on fraction $x$. However, $a$ is not a constant, it depends on $\tau$. Therefore, the actual reach at HL-LHC for a particular process ($\tau_1$) affects the needed luminosity. Hence, treating $a$ as a constant at most only conveys the qualitative feature of the scaling. In addition, the parton luminosity has a logarithmic dependence on the overall energy scale, which we have ignored in this scaling argument (although it has been taken into account in the numerical result presented in \autoref{fig:ppLumiE}).

% \clearpage
% \subsection{Other BSM scenarios}
% Placeholder for other motivated cases that we haven't discussed so far.

% \section{Summary and Outlook}

% Implementation frontier. Accelerator frontier. Energy Frontier. Physics. 

\bibliographystyle{utphys}
\typeout{}
\bibliography{references}

\end{document}